\documentclass[letterpaper]{article} 
\usepackage[preprint]{aaai2027}  
\usepackage[hyphens]{url}  
\usepackage{graphicx} 
\def\UrlFont{\rm}  
\usepackage{natbib}  
\usepackage{caption} 
\usepackage{booktabs}
\usepackage{multirow}
\usepackage{amsmath}
\usepackage{amssymb}
\newcommand{\pmerr}[1]{_{\scriptscriptstyle\pm #1}}

\title{VocalRender: Score-Native Singing Voice Synthesis for Real-World Composition}

\author{
    Yukun Chen\textsuperscript{\rm 1,\rm 2},
    Tianrui Wang\textsuperscript{\rm 2},
    Zhaoxi Mu\textsuperscript{\rm 3,\rm 4},
    Xinyu Yang\textsuperscript{\rm 1},
    EngSiong Chng\textsuperscript{\rm 2}
}
\affiliations{
    \textsuperscript{\rm 1}Xi'an Jiaotong University
    \quad
    \textsuperscript{\rm 2}Nanyang Technological University\\
    \textsuperscript{\rm 3}Ant Group
    \quad
    \textsuperscript{\rm 4}Zhejiang University\\
    chenyk@stu.xjtu.edu.cn
}

\begin{document}

\maketitle

\begin{abstract}

Existing singing voice synthesis systems often require predefined durations, explicit duration prediction, or time-aligned acoustic guidance, which limits their compatibility with practical composition workflows. We propose VocalRender, a score-native system that directly synthesizes singing from lyrics, pitches, symbolic note values, and tempo. It uses an interleaved lyric–note representation and an autoregressive diffusion model to generate continuous acoustic latents while predicting the output length, eliminating the need for explicit duration prediction. Trained on a 2,300-hour singing dataset, VocalRender achieves strong intelligibility, strong melody control, and high speaker similarity across both in-domain and out-of-domain benchmarks. Notably, it outperforms the strongest baseline by $0.42$ points in naturalness CMOS, demonstrating the effectiveness of our proposed score-native architecture.

\end{abstract}

\begin{links}
    \link{Code}{https://github.com/pymaster17/VocalRender}
\end{links}

\section{Introduction}

\begin{figure*}[t]
\centering
\includegraphics[width=0.8\textwidth]{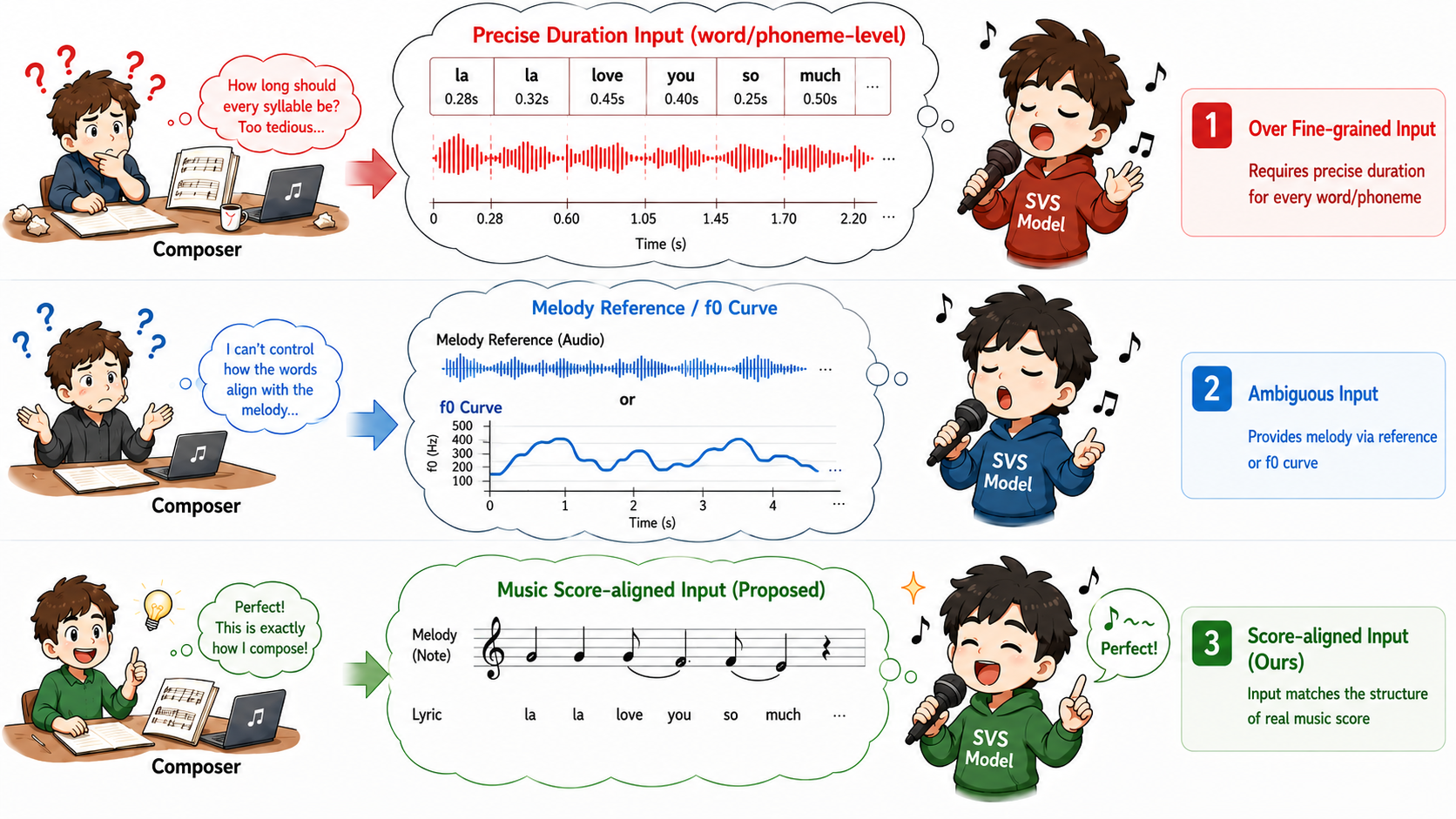}
\caption{SVS with different input styles. \textbf{Top:} duration-based SVS \textbf{Middle:} reference-based SVS \textbf{Bottom:} score-native SVS}
\label{fig:intro}
\end{figure*}

Singing voice synthesis (SVS) aims to produce singing audio with similar performance quality to human singers. The task is highly relevant to text-to-speech (TTS), both of which generate human voice conditioning on text-style input. However, two main distinctions remain at input and output ends, respectively. Apart from text, SVS models need additional constraints like pitch and note to ensure the generated audio follows the desired melody. Although both are human voice, singing is far more expressive than speech, exhibiting larger variations in pitch and prosody.

The development path of SVS highly imitates TTS, where the early models \citep{ren2020deepsinger,liu2022diffsinger,wang2022xiaoicesing,zhang2024tcsinger,he2023rmssinger,guo2025techsinger,zhang2022visinger,zhang2022visinger2} follow the main architecture of successful TTS models \citep{ren2020fastspeech,kim2021conditional}, with extended input formats and improved F0 modeling. With audio codecs being widely adopted in TTS models, several attempts on SVS task \citep{hwang2025hiddensinger,wu2024toksing} are also made for their higher representation efficiency. These models either require direct duration input, or an explicit duration predictor. Although the architecture can achieve high controllability and acceptable naturalness with limited training data, a unified architecture without direct duration input and explicit duration predictor is needed for more composer-friendly and higher performance ceiling. Moreover, the requirement of fine-grained annotations largely prevents them from scaling up.

Recently, Diffusion Transformer (DiT) has been explored in SVS \citep{du2026ditsinger,zheng2025yingmusic,qian2026soulx}, for better long-range consistency, flexible condition injection and better scalability. Although more unified, DiT module needs a predefined duration for denoising, which means either a time-aligned cue \citep{zheng2025yingmusic,qian2026soulx} or a duration predictor frontend is needed \citep{zhang2025tcsinger,du2026ditsinger}. Moreover, most of these models treat lyric and melody conditions as two independent streams, which can cause an ambiguous alignment between lyrics and pitches when melisma exists.

Scaling models tend to choose acoustic melody cues like reference audio or F0 curves \citep{zhang2025vevo2,zheng2025yingmusic,qian2026soulx}, as symbolic and music theory consistent annotations are difficult to scale with audio. Even ignoring the annotation rareness, the overall scale of existing singing audio datasets \citep{wang2022opencpop,zhang2022m4singer,zhang2024gtsinger,huang2021multi} lags far behind its TTS counterparts, which have been extended to 100k hours \citep{he2024emilia}. The insufficient data prevent SVS systems from adopting modern architectures which rely on data scaling for a higher performance ceiling \citep{zhang2025vevo2,shi2024singing,gu2025singnet}. Although attempts have been made to collect in-the-wild singing data \citep{ren2020deepsinger,gu2025singnet}, relying on web-crawled audio often suffers from acoustic interference and the inherent inaccuracies of weak annotations, severely limiting the reliability of the learned alignments.

To eliminate the need for an explicit duration predictor, we introduce the autoregressive diffusion model (ARDM) \citep{jia2025ditar} into SVS, in which the AR module can generate a prosody sketch without time-aligned signals, and the LocDiT module can further generate high-fidelity audio latents. To keep the input format equivalent to real music scores, we develop an interleaved prompt structure, with the pitch and note tokens strictly tuned according to music theory, as illustrated in Figure~\ref{fig:intro}. To effectively model the complex mapping between such symbolic scores and highly expressive acoustics, our ARDM framework requires a strong foundation of reliable, fine-grained annotations. Therefore, we utilize a massive 2,300-hour dataset meticulously extracted and refined from diverse open-source collections. This robust data foundation—over 20$\times$ larger than existing ones—provides the essential alignment precision and acoustic diversity required by our model. The main contributions of this paper are summarized as follows:
\begin{itemize}
    \item {We introduce VocalRender, the first ARDM-based model for the SVS task.} By eliminating the explicit duration predictor, it demonstrates impressive scalability and achieves competitive performance beyond previous architectures.
    \item {We develop a composer-friendly, interleaved input representation.} This structure is intrinsically compatible with symbolic music scores, effectively solving the indeterministic alignment problem between lyrics and pitches when melisma exists.
    \item {We demonstrate state-of-the-art generation capabilities on Mandarin.} Empowered by large-scale, fine-grained annotated data, VocalRender achieves highly natural and expressive singing synthesis, validating the effectiveness of the explicit duration-free modeling paradigm.
\end{itemize}

\begin{figure*}[t]
\centering
\includegraphics[width=0.9\textwidth]{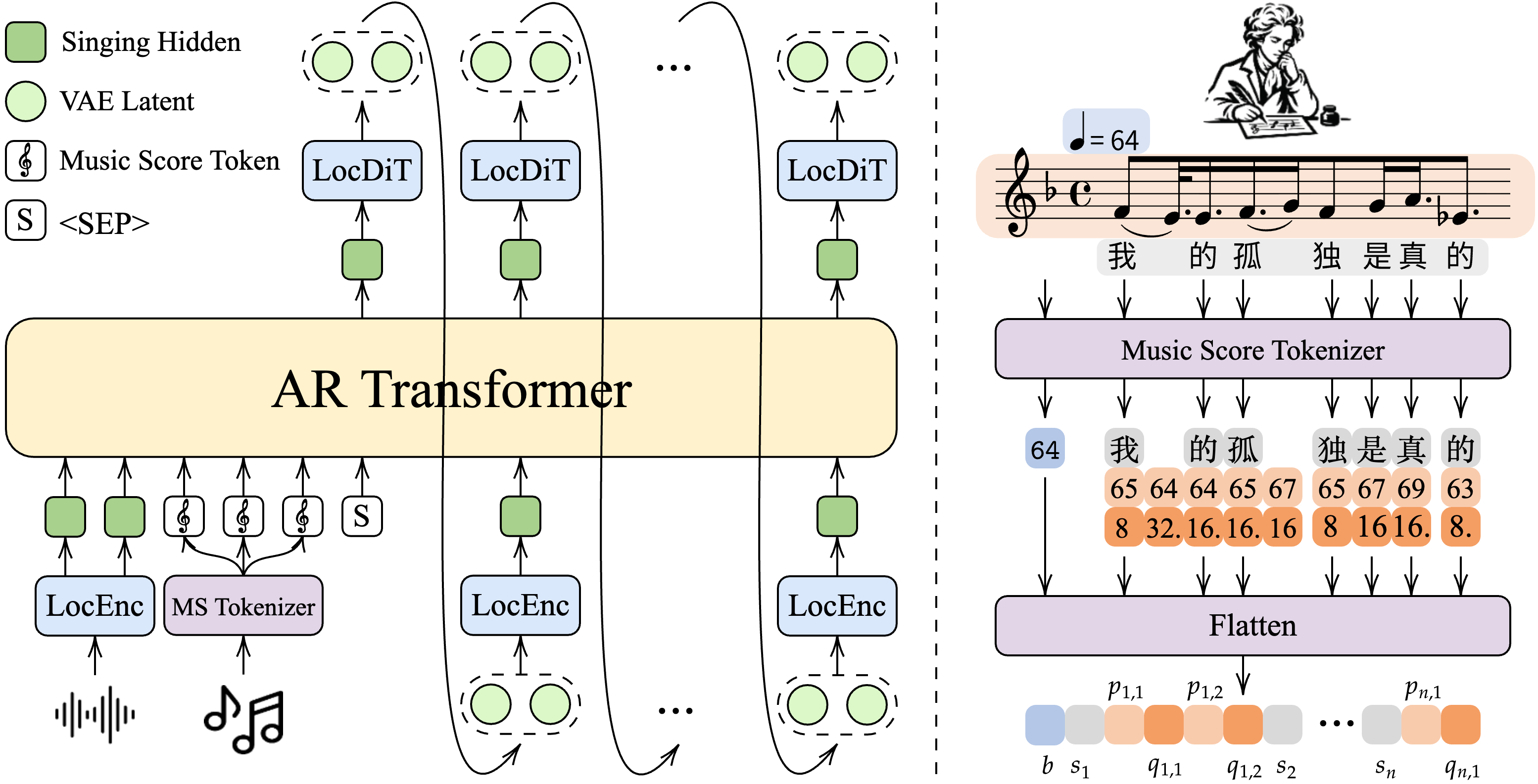}
\caption{Overall structure of VocalRender. \textbf{Left:} the autoregressive diffusion framework. \textbf{Right:} the music score tokenization process}
\label{fig:structure}
\end{figure*}

\section{Related Works}

\subsection{Singing Voice Synthesis}

The early models of SVS characterized by strong guidance inputs and explicit F0 and duration prediction, as they are how singing differs from plain speech. DeepSinger \citep{ren2020deepsinger} and XiaoiceSing \citep{lu2020xiaoicesing} adopt feed-forward transformers for mel-spectrogram generation, followed by a vocoder for decoding waveform. DiffSinger \citep{liu2022diffsinger} improves the frontier step by conditional diffusion for more stable training and better audio quality. VISinger2 \citep{zhang2022visinger2} first introduces an end-to-end architecture based on VITS, with a DDSP-integrated Mel decoder to reduce artefacts. RMSSinger \citep{he2023rmssinger} simplifies the duration inputs from phoneme-level to word-level, making the composition process more convenient and improving the singing naturalness. TechSinger \citep{guo2025techsinger} further improves the acoustic modeling with flow-matching, with controllability for multiple vocal techniques. TCSinger2 \citep{zhang2025tcsinger} focuses on style transfer through a contrastive learning based audio encoder.

Recent works turn to more unified DiT framework for better long-range consistency, flexible conditions injection and better scalability. Meanwhile, weaker guidance generation is widely adopted, liberating data collection from labour-intensive annotation processes. YingMusic-Singer \citep{zheng2025yingmusic} aligns a melody extractor with teacher model during the end-to-end training, which can extract melody of the reference audio and transfer it into target audio. SoulX-Singer \citep{qian2026soulx} adopts either note pitch sequence or F0 curve as melody cues, where the note types are simplified to 3 categories:
rest, lyric, and slur. Vevo2 \citep{zhang2025vevo2} builds on a cascaded structure of AR and NAR modules, responsible for semantic and acoustic modeling. A prosody tokenizer trained on chromagram reconstruction is used to extract prosody cues from reference audio. The input formats of these models are either too trivial to compose, or too ambiguous to produce desired prosody robustly, preventing their deployment in the serious music creation scenario.

\subsection{Music data scaling}

As the infrastructure of SVS training, the quality and scale of music data is deterministic for the performance. Opencpop \citep{wang2022opencpop} is a single-singer singing dataset with fine-grained phoneme-level annotations . M4Singer \citep{zhang2022m4singer} extended to multiple singers with about 30 hours in total. GTSinger \citep{zhang2024gtsinger} further includes multiple languages with additional vocal techniques labels. Although they have high-quality singing records and high reliable annotations under professional inspection, none of them have total duration exceeds 100 hours. In contrast, as reported in the recent works \citep{du2026ditsinger,zheng2025yingmusic,zhang2025vevo2,qian2026soulx}, the data demand for modern transformer based models ranges from 500 hours to 42k hours. Although attempts are made for large scale singing dataset \citep{gu2025singnet,wang2025multi}, none of them are publicly available at our best knowledge.

Meanwhile, the full song dataset emerged in the recent works. MuChin \citep{wang2024muchin} collect over 6000 songs with rich metadata at multiple aspects. SongFormDB \citep{hao2025songformer} contains about 6000 songs with well-aligned structural timestamps. Muse \citep{jiang2026muse} consists of 116k fully licensed synthetic songs, using GPT-5 as composer and SunoV5 as singer. However, these audios are vocal/accompany mixture, and lack music-score annotations, so can not be used for SVS training directly.

\section{VocalRender}

\subsection{Overview}

VocalRender is a singing voice synthesis system that generates realistic singing audio from composer-oriented symbolic scores, whose overall architecture is illustrated in Figure~\ref{fig:structure}. The system is built upon three key designs. First, a score-native representation serializes lyrics, MIDI pitches, and symbolic note values into a syllable-level interleaved sequence, explicitly preserving the one-to-many correspondence between lyrics and notes. Second, a continuous audio variational autoencoder (VAE) encodes the target waveform into a compact latent sequence, avoiding the information bottleneck introduced by discrete acoustic tokenization. Third, an autoregressive diffusion model (ARDM) generates the acoustic latent sequence patch by patch and predicts when the generation should terminate. This formulation allows VocalRender to jointly determine local timing and total output duration during generation, without requiring phoneme-level duration labels, an explicit duration predictor, or a time-aligned acoustic reference. The generated latent sequence is finally converted into a singing waveform by the VAE decoder.

\subsection{Task Formulation}
\label{sec:task_formulation}

Let a symbolic music score be represented as
\begin{equation}
\mathcal{S}
=
\left(
b,
\left\{
\left(s_i,\mathcal{A}_i\right)
\right\}_{i=1}^{N}
\right),
\end{equation}
where $b$ denotes the global beats per minute (BPM), $s_i$ is the $i$-th lyric syllable, and $\mathcal{A}_i$ contains the notes assigned to that syllable. Specifically,
\begin{equation}
\mathcal{A}_i
=
\left[
\left(p_{i,1},q_{i,1}\right),
\ldots,
\left(p_{i,K_i},q_{i,K_i}\right)
\right],
\end{equation}
where $p_{i,k}$ and $q_{i,k}$ denote the MIDI pitch and symbolic note value of the $k$-th note associated with $s_i$, respectively. Here, a note value denotes a composer-authored rhythmic category, such as a quarter note, eighth note, or dotted note, rather than a realized duration in seconds. $K_i=1$ corresponds to the conventional one-syllable--one-note case, whereas $K_i>1$ represents melisma, in which a syllable spans multiple notes.

Given an audio representation $z \in \mathbb{R}^{T\times d}$, where $T$ is the latent frame length and $d$ is the latent dimension, the objective of score-conditioned singing voice synthesis is therefore to model
\begin{equation}
p_{\theta}
\left(
\mathbf{z}
\mid
\mathcal{S}, \mathbf{r}
\right),
\end{equation}
where $\mathbf{r}$ denotes an optional reference-audio prompt used for timbre conditioning.

A central challenge is that $T$ is unknown at inference time. Symbolic note values describe nominal rhythmic relations rather than exact acoustic durations: the realized duration of each syllable and note can vary with articulation, melisma, breathing, and other expressive choices. Consequently, the model must infer both the acoustic realization and its sequence length directly from the symbolic score. However, conventional non-autoregressive models operate on a predefined acoustic canvas and therefore require the output sequence length before generation begins. To obtain this length, existing systems typically rely on either explicit duration predictors---which rigidly commit to fine-grained timing before acoustic generation---or time-aligned acoustic cues (e.g., reference audio or F0 contours)---which may not be available during composition and constrain the generated prosody. This highlights the need for a framework capable of jointly determining local timing and total duration dynamically.

\subsection{Representation}

\subsubsection{Audio Representation}

Continuous generative models based on diffusion or flow matching are well suited to high-fidelity acoustic modeling, provided they operate on an appropriate feature space. VocalRender models singing audio in a learned continuous latent space. Unlike discrete acoustic tokenizers, which may discard fine-grained pitch, timbre, and articulation details during quantization, continuous latents preserve the exact acoustic information needed for highly expressive singing generation. Compared with mel-spectrograms, learned latents are also more compact and reduce the sequence length seen by the acoustic generator.

Given a waveform $\mathbf{x}$, we use an audio variational autoencoder (VAE) with encoder $\mathcal{E}$ and decoder $\mathcal{D}$. The encoder maps the waveform into a latent sequence
\begin{equation}
\mathbf{z}
=
\mathcal{E}(\mathbf{x}).
\end{equation}
In our implementation, the VAE is trained before the ARDM and then kept fixed during score-conditioned generation training. The latent frame rate is 25 Hz, which provides a compact acoustic canvas while retaining sufficient resolution for local singing dynamics. This design avoids discrete acoustic tokenization and allows the diffusion module to operate directly on continuous acoustic variables.

\subsubsection{Music Score Representation}

The symbolic score in Section~\ref{sec:task_formulation} has a nested structure: each lyric syllable $s_i$ is associated with a note list $\mathcal{A}_i=[(p_{i,1},q_{i,1}),\ldots,(p_{i,K_i},q_{i,K_i})]$. The global BPM $b$ provides the tempo reference that converts these relative note values into a nominal rhythmic scale, while the model remains free to realize expressive deviations during generation.

To use the score as input to a decoder-only transformer, we flatten the nested syllable--note structure into a one-dimensional token sequence while preserving local correspondence. For the $i$-th syllable, its associated note list is serialized as
\begin{equation}
\mathbf{a}_i
=
\bigoplus_{k=1}^{K_i}
\left(
p_{i,k}
\oplus
q_{i,k}
\right),
\end{equation}
where $\oplus$ denotes sequence concatenation. The complete score prompt is then
\begin{equation}
\mathbf{c}(\mathcal{S})
=
b
\bigoplus_{i=1}^{N}
\left(
s_i
\oplus
\mathbf{a}_i
\right).
\end{equation}

This interleaved serialization explicitly binds each syllable to the notes on which it should be sung. When $K_i=1$, the sequence reduces to the conventional one-syllable--one-note case. When $K_i>1$, the same syllable is followed by multiple pitch--note-value pairs,
\begin{equation}
s_i
\oplus
(p_{i,1}\oplus q_{i,1})
\oplus
\cdots
\oplus
(p_{i,K_i}\oplus q_{i,K_i}),
\end{equation}
which naturally represents melisma without requiring phoneme-level alignment or manually assigned note durations.

\subsection{Architecture}

\begin{table*}[t]
\centering
\small
\setlength{\tabcolsep}{4pt}
\begin{tabular}{lcccccc@{\hspace{1.8em}}cccccc}
\toprule
\multirow{2}{*}{Method} & \multicolumn{6}{c}{\textbf{Opencpop}} & \multicolumn{6}{c}{\textbf{CrawlSinger-Eval}} \\
\cmidrule(lr){2-7}\cmidrule(lr){8-13}
& WER$\downarrow$ & SIM$\uparrow$ & IOU$\uparrow$ & RPA$\uparrow$ & SingMOS$\uparrow$ & CE$\uparrow$ & WER$\downarrow$ & SIM$\uparrow$ & IOU$\uparrow$ & RPA$\uparrow$ & SingMOS$\uparrow$ & CE$\uparrow$ \\
\midrule
GT & $3.81$ & -- & -- & -- & $4.59$ & $5.61$ & $5.17$ & -- & -- & -- & $4.49$ & $5.73$ \\
\midrule
TCSinger & $21.67$ & $0.891$ & $0.59$ & $0.67$ & $3.91$ & $5.65$ & $26.13$ & $0.853$ & $0.46$ & $0.56$ & $3.84$ & $5.29$ \\
TechSinger & $14.55$ & -- & $0.51$ & $0.69$ & $4.21$ & $5.32$ & $13.72$ & -- & $0.42$ & $0.59$ & $4.16$ & $5.35$ \\
Vevo2 & $14.42$ & $0.866$ & $0.27$ & $0.60$ & $4.12$ & $\underline{5.85}$ & $27.27$ & $0.874$ & $0.16$ & $0.56$ & $4.08$ & $5.75$ \\
SoulX-Singer & $5.02$ & $\underline{0.928}$ & $\mathbf{0.63}$ & $\mathbf{0.77}$ & $4.45$ & $\mathbf{6.01}$ & $5.03$ & $0.918$ & $\mathbf{0.54}$ & $\mathbf{0.70}$ & $4.44$ & $5.71$ \\
\midrule
VocalRender & $\underline{4.44}$ & $0.922$ & $\underline{0.62}$ & $0.72$ & $\mathbf{4.59}$ & $\underline{5.85}$ & $\underline{4.52}$ & $\underline{0.919}$ & $0.43$ & $0.63$ & $\mathbf{4.53}$ & $\mathbf{5.85}$ \\
VocalRender-Pro & $\mathbf{3.88}$ & $\mathbf{0.929}$ & $\underline{0.62}$ & $\underline{0.75}$ & $\underline{4.55}$ & $5.63$ & $\mathbf{4.45}$ & $\mathbf{0.926}$ & $\underline{0.48}$ & $\underline{0.67}$ & $\underline{4.52}$ & $\underline{5.78}$ \\
\bottomrule
\end{tabular}
\caption{Performance of different model on Opencpop and CrawlSinger-Eval.}
\label{tab:svs_performance}
\end{table*}

\begin{table}[t]
\centering
\begin{tabular}{lccc}
\toprule
Method & N-CMOS & PS-CMOS & MS-MOS \\
\midrule
TCSinger & $-1.20\pmerr{0.26}$ & $-1.53\pmerr{0.19}$ & $2.19\pmerr{0.22}$ \\
TechSinger & $-0.99\pmerr{0.26}$ & $-1.16\pmerr{0.25}$ & $2.56\pmerr{0.28}$ \\
Vevo2 & $-0.93\pmerr{0.29}$ & $-1.65\pmerr{0.18}$ & $2.12\pmerr{0.27}$ \\
SoulX-Singer & $-0.42\pmerr{0.28}$ & $-0.32\pmerr{0.24}$ & $\underline{2.84}\pmerr{0.22}$ \\
\midrule
VocalRender & $\underline{0}$ & $\underline{0}$ & $\mathbf{2.96}\pmerr{0.22}$ \\
VocalRender-Pro & $\mathbf{+0.13}\pmerr{0.27}$ & $\mathbf{+0.06}\pmerr{0.22}$ & $2.71\pmerr{0.27}$ \\
\bottomrule
\end{tabular}
\caption{Subjective evaluation results of different models.}
\label{tab:subjective_evaluation}
\end{table}

\begin{table*}[t]
\centering
\small
\setlength{\tabcolsep}{4pt}
\begin{tabular}{lcccccc@{\hspace{1.8em}}cccccc}
\toprule
\multirow{2}{*}{Setting} & \multicolumn{6}{c}{\textbf{Opencpop}} & \multicolumn{6}{c}{\textbf{CrawlSinger-Eval}} \\
\cmidrule(lr){2-7}\cmidrule(lr){8-13}
& WER$\downarrow$ & SIM$\uparrow$ & IOU$\uparrow$ & RPA$\uparrow$ & SingMOS$\uparrow$ & CE$\uparrow$ & WER$\downarrow$ & SIM$\uparrow$ & IOU$\uparrow$ & RPA$\uparrow$ & SingMOS$\uparrow$ & CE$\uparrow$ \\
\midrule
Base & $4.44$ & $0.922$ & $0.62$ & $0.72$ & $4.59$ & $5.85$ & $4.52$ & $0.919$ & $0.43$ & $0.63$ & $4.53$ & $5.85$ \\
\midrule
w/o CrawlSinger-OS & $+0.12$ & $-0.031$ & $-0.13$ & $-0.45$ & $-0.04$ & $+0.06$ & $+0.03$ & $-0.035$ & $-0.09$ & $-0.40$ & $+0.06$ & $-0.04$ \\
w/o interleaving & $+0.11$ & $+0.001$ & $-0.05$ & $+0.02$ & $-0.01$ & $-0.01$ & $+0.16$ & $+0.001$ & $-0.04$ & $+0.01$ & $-0.02$ & $-0.02$ \\
\bottomrule
\end{tabular}
\caption{Ablation study results on Opencpop and CrawlSinger-Eval.}
\label{tab:ablation_study}
\end{table*}

To avoid both the predefined frame allocation of duration-based models and the reliance on time-aligned acoustic guidance, VocalRender adopts an autoregressive diffusion formulation. This design combines the flexible length and sequential planning capabilities of autoregressive generation with the high-fidelity continuous modeling capability of diffusion, enabling natural timing to emerge directly from score-native inputs. 
Specifically, VocalRender uses this ARDM to synthesize continuous singing latents directly from a symbolic music-score prompt. We divide the target continuous latent $\mathbf{z}$ into $Y$ local patches:


\begin{equation}
\mathbf{Z} = [\mathbf{z}^{(1)}, \mathbf{z}^{(2)}, \dots, \mathbf{z}^{(Y)}],
\quad
\mathbf{z}^{(y)} \in \mathbb{R}^{P \times d},
\end{equation}
where $P$ denotes the patch size and $Y=\lceil T/P\rceil$ during training. Instead of requiring a predefined acoustic length, ARDM factorizes the conditional distribution autoregressively over latent patches:

\begin{equation}
p_{\theta}(\mathbf{Z} \mid \mathcal{C})
=
\prod_{y=1}^{Y}
p_{\theta}\left(
\mathbf{z}^{(y)}
\mid
\mathcal{C}, \mathbf{z}^{(<y)}
\right).
\end{equation}

This formulation allows the model with parameter $\theta$ to generate singing audio progressively, rather than being constrained by fixed prosody from external duration predictor. For reference-based timbre conditioning, we prepend continuous latent patches extracted from a short audio prompt to the music-score tokens. Following common in-context learning practice in autoregressive audio generation, the model conditions on these prompt latents as acoustic context while generating the target singing sequence.

The ARDM model can be divided into AR transformer with parameter $\theta_a$ and DiT with parameter $\theta_b$. For each latent patch, an aggregation encoder maps the local continuous representation into a compact embedding:

\begin{equation}
\mathbf{e}_y = \mathrm{Agg}_{\psi}\left(\mathbf{z}^{(y)}\right).
\end{equation}

which is fed into AR transformer as acoustic history context. Then, the AR transformer outputs conditioning on symbolic score and acoustic history:
\begin{equation}
\mathbf{h}_y =
\mathrm{AR}_{\theta_a}
\left(
\mathcal{C}, \mathbf{e}_{<y}
\right),
\end{equation}
where $\mathbf{h}_y$ serves as the condition of DiT for generating the next latent patch, together with the last acoustic history patch
\begin{equation}
\mathbf{z}^{(y)} =
\mathrm{DiT}_{\theta_b}
\left(
\mathbf{h}_y, \mathbf{z}^{(y-1)}
\right),
\end{equation}
The whole ARDM model is trained end-to-end by a simple flow-matching loss on acoustic latent space

\begin{equation}
\begin{aligned}
\mathcal{L}_{\mathrm{FM}} = & \mathbb{E}_{y,t,\boldsymbol{\epsilon }}[\Vert \mathbf{v}_{\theta _{b}}\left(\mathbf{z}_{t}^{(y)} ,t,\mathbf{h}_{y} ,\mathbf{z}_{0}^{(y-1)}\right)\\
 & -\frac{d}{dt}\left( \alpha _{t}\mathbf{z}_{0}^{(y)} +\sigma _{t}\boldsymbol{\epsilon }\right)\Vert _{2}^{2}] .
\end{aligned}
\end{equation}
A binary stop predictor conditioning on AR hidden $h_y$ is trained jointly, outputting stop flag when the generation process completes. The final loss function of ARDM is $\mathcal{L}_{\mathrm{ARDM}} = \mathcal{L}_{\mathrm{FM}} + \lambda_{stop}\mathcal{L}_{stop}$, where $\lambda_{stop}=1$ is adopted in practice.

\section{Experiments}

\subsection{Experimental Setup}

\subsubsection{Dataset}

To utilize the scalability of ARDM architecture, we collect two large scale singing datasets: CrawlSinger and CrawlSinger-OS, following SingCrawl pipeline \citep{chen2026vocalparse}. CrawlSinger has over 5600 hours of high-quality singing audio extracted from in-house song data, paired with syllable-level pitch and duration labels. CrawlSinger-OS shares the same processing pipeline and annotation scheme with CrawlSinger, but is constructed from multiple publicly available song and singing datasets, including OpenSinger, Muchin, SongFormDB and Muse, with a total duration of over 2300 hours. Further details on CrawlSinger-OS are provided in Appendix~B.

We use Opencpop as the evaluation dataset. To avoid data leakage from web-crawled dataset, we filter out part of songs in Opencpop based on lyric similarity. To further evaluate the out-of-domain (OOD) performance, we also construct CrawlSinger-Eval (2606), including newly created songs with published time after 2026-06-01, later than all the songs in the training set. We deliberately choose original songs with limited popularity for CrawlSinger-Eval, avoiding potential style and melody duplication.

\subsubsection{Implementation Details}

We adopt VoxCPM2 \citep{zhou2026voxcpm2} as the ARDM backbone and initialize from speech-pretrained weights. VocalRender and VocalRender-Pro are trained on CrawlSinger-OS and CrawlSinger, respectively. For reference-conditioned generation, a randomly sampled 2-8s segment from the same song is used as the acoustic prompt. VocalRender-Pro is trained for 160k steps with a global batch size of 32,768 tokens, while VocalRender follows a two-stage training strategy with synthetic pretraining and realistic-data finetuning. Additional architectural and optimization details are provided in Appendix~C.

\subsubsection{Evaluation Metrics}

For objective evaluation, we use WER for intelligibility and SIM for speaker similarity. For WER, we employ Qwen3-ASR \citep{shi2026qwen3} for its robust performance on singing audio. For SIM, we compute the cosine similarity of speaker embeddings extracted by WavLM TDNN \citep{chen2022wavlm}, following Vevo2 \citep{zhang2025vevo2}. The Intersection Over Union (IOU) score between word alignment and Raw Pitch Accuracy (RPA) between pitch alignment are used for rhythm and melody similarity. The alignment is transcribed end-to-end by STARS \citep{guo2025stars}. We also introduce two learning-based perceptual metrics. Specifically, SingMOS \citep{tang2024singmos} for singing-specific quality evaluation and CE from Audiobox Aesthetics \citep{tjandra2025meta}, which is reported highly relevant with vocal quality in YuE \citep{yuan2025yue}.

For subjective evaluation, we employ Comparative Mean Opinion Score (CMOS) to evaluate naturalness and prosody similarity with N-CMOS and PS-CMOS, respectively. CMOS rates from -2 to 2, representing one sample is much worse/better than the other. Moreover, we use MS-MOS to evaluate the music score following ability, rating from 1 to 4 or from unable to follow to strictly follow. Detailed subjective-evaluation settings are provided in Appendix~A.

\subsubsection{Baseline Models}

We compare VocalRender and VocalRender-Pro with four frontier SVS models: TCSinger \citep{zhang2024tcsinger}, a zero-shot singing model with good style transfer ability; TechSinger \citep{guo2025techsinger}, a flow-matching singing model with technique controllability; Vevo2 \citep{zhang2025vevo2}, a unified speech and singing with cascaded AR transformer and DiT; SoulX-Singer \citep{qian2026soulx}, a DiT singing model with large scale training. Because the evaluated systems accept different conditioning formats, we construct all baseline inputs from the same information available in a conventional symbolic score. For models requiring phoneme- or note-level durations, we convert each symbolic note value into a nominal physical duration according to the score BPM, rather than using durations aligned to the ground-truth performance. This setting is intended to evaluate practical score-conditioned synthesis under an equal information budget.

\subsection{Main Results}

\subsubsection{Intelligibility and Timbre Similarity}
VocalRender achieves the lowest Word Error Rate (WER) across all baselines. It reaches a WER of $4.44$ on Opencpop and $4.52$ on the CrawlSinger-Eval. These values are close to or even surpass the ASR-based WERs measured on the ground-truth recordings, indicating that the generated lyrics are highly intelligible. The slight difference can be explained by occasionally slurred lyrics for stylistic purposes. VocalRender also shows top-tier timbre similarity of $0.922$ and $0.919$ on both eval datasets through simple in-context learning, competitive with the strongest baseline SoulX-Singer.

\subsubsection{Rhythm and Melody Control}
Without relying on an explicit duration predictor or time-aligned acoustic reference, VocalRender successfully demonstrates strong controllability over rhythm and melody. While SoulX-Singer achieves the highest objective rhythm (IOU) and pitch (RPA) scores due to its reliance on strict predefined temporal alignments, subjective evaluations reveal the perceptual advantage of our approach. In the Music Score Following test (MS-MOS), VocalRender achieves the highest score of $2.96$, indicating superior subjective adherence to the composer's intent. Furthermore, the Prosody Similarity CMOS (PS-CMOS) demonstrates that our models significantly outperform all baselines, including SoulX-Singer ($-0.32$), the winner in objective test. This confirms that our proposed interleaved prompt format effectively guides the AR module to generate accurate, natural prosody sketches strictly from symbolic music scores, avoiding the mechanical rigidity often caused by traditional length regulators.

\subsubsection{Musical Quality}
VocalRender obtains SingMOS scores of $4.59$ on Opencpop and $4.53$ on CrawlSinger-Eval, which are comparable to the scores assigned by the automatic metric to the ground-truth recordings. Several recent systems, including SoulX-Singer and VocalRender-Pro, also receive closely clustered SingMOS and CE scores. The small differences suggest that the evaluated automatic perceptual metrics may have limited discriminative resolution among frontier generation models.

However, the subjective N-CMOS test shows that the musical quality difference among those models are still obvious. VocalRender substantially surpasses all baseline models, with the most competitive baseline (SoulX-Singer) lags behind at $-0.42$. Informal feedback from professional listeners suggests that the perceived differences may be related in part to the realization of melismatic pronunciation. We believe the advantage owes to the AR transformer of VocalRender, which is generally better at semantic sketch modeling than NAR counterparts.

\subsubsection{Out-of-Domain Generalization}
Comparing with Opencpop, CrawlSinger-Eval includes more fresh songs collected later than training data, which is designed for model generalization ability. TCSinger and Vevo2 exhibit larger degradations in intelligibility, with WER jumping to $>26$, and TCSinger faces further degradation on speaker similarity. In contrast, VocalRender, VocalRender-Pro, and SoulX-Singer maintain relatively stable performance on most metrics, indicating that the VocalRender models generalize effectively to unseen lyric, timbre and singing style. Furthermore, we observe that models trained on larger dataset (VocalRender-Pro, SoulX-Singer) are likely to be more robust under OOD scenarios, with moderate decreases on IOU and RPA.

\subsubsection{Influence of Training Data}
VocalRender and VocalRender-Pro share the same architecture but differ in training data and strategy. As shown in Table, VocalRender-Pro consistently achieves better intelligibility and speaker similarity. On Opencpop, it reduces WER from $4.44$ to $3.88$ and improves SIM from $0.922$ to $0.929$. Similar improvements are observed on CrawlSinger-Eval, where WER decreases from $4.52$ to $4.45$ and SIM increases from $0.919$ to $0.926$. These results suggest that the substantially larger amount of real singing data and the broader singer coverage benefit both lyric pronunciation modeling and singer identity preservation. VocalRender-Pro also obtains positive N-CMOS and PS-CMOS scores of $+0.13$ and $+0.06$, respectively, when VocalRender is used as the reference. This indicates that its generated samples are perceived as slightly more natural and stylistically closer to real singer performances. However, VocalRender achieves a higher MS-MOS score of $2.96$, compared with $2.71$ for VocalRender-Pro. This result indicates that VocalRender follows the input music score more accurately despite being trained on less real data. A possible explanation is that its finetuning subset contains more reliable and precise score annotations.

\section{Ablation Study}

We conduct ablation experiments to investigate the contributions of the large-scale CrawlSinger-OS training data and the proposed lyric--note interleaved input format. Table~\ref{tab:ablation_study} reports metric changes relative to the complete VocalRender model, where positive and negative values denote increases and decreases, respectively.

\subsection{Effect of CrawlSinger-OS}

We remove CrawlSinger-OS and train VocalRender only on GTSinger and M4Singer datasets. The results show that the WER only increases marginally, with $+0.12$ and $+0.03$ on Opencpop and CrawlSinger-Eval. This may be attributed to the strong intelligibility bias of the speech-pretrained backbone. However, the degradation of speaker similarity is obvious, which drops over $0.03$. It is worth noticing that the performance is highly consistent with TCSinger, which was trained with a similar data scale, proving that large-scale and diversity data is helpful for singing voice cloning. Furthermore, data limitation is harmful for prosody modeling, especially for the melody, where the RPA decreases more than $0.4$. It proves that the model is still highly underfit on the newly added music tokens when the training size is limited.

\subsection{Effect of Interleaved Prompting}

We also replace the proposed interleaved prompt with cascaded style, where the musical attribute sequence $A$ is cascaded after the complete lyric sequence $S$. The style change disrupts the alignment between syllables and musical attributes, without influencing other information. The results prove the interleaved prompt is effective to solve the rhythm ambiguity when providing syllable--note alignment implicitly. When lacking the alignment information, the rhythm similarity degrades significantly, although melody is nearly unaffected. 

\section{Discussion}

\begin{figure}[t]
\centering
\includegraphics[width=\columnwidth]{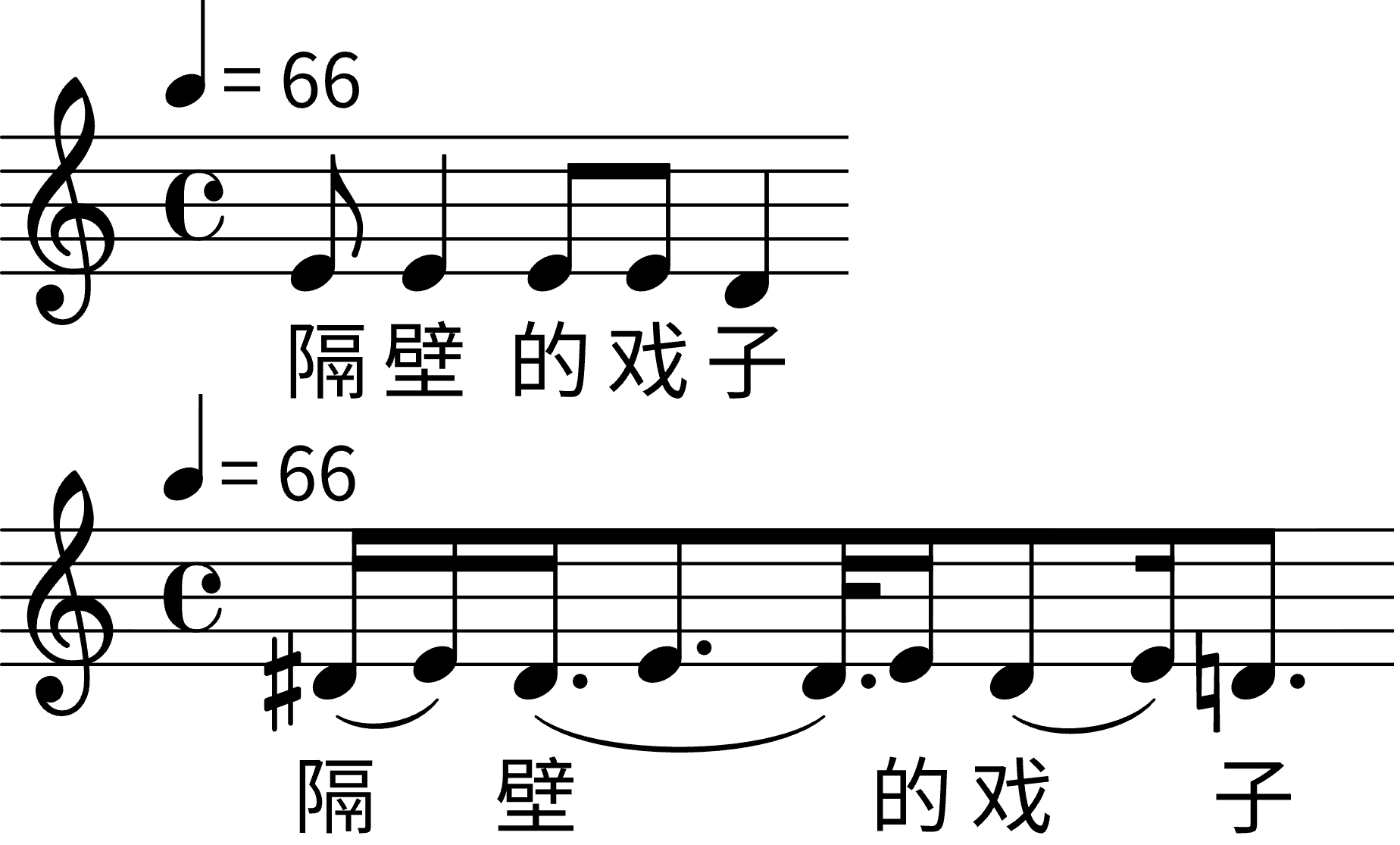}
\caption{Comparison between real music score and transcribed music score. Upper: \emph{intent-centric} music score. Bottom: \emph{audio-centric} music score}
\label{fig:GT_vs_Auto}
\end{figure}

A potential limitation of our training pipeline is the mismatch between composer-authored and automatically transcribed scores. Human-written scores are primarily \emph{intent-centric}, specifying the main melodic and rhythmic structure while leaving expressive realization to the singer. For example, a singer may introduce additional melismas, note splitting, or ornamental pitches that are not explicitly written in the original score. In contrast, an \emph{audio-centric} transcription attempts to recover every realized note and therefore produces a substantially more detailed condition (Figure~\ref{fig:GT_vs_Auto}).

A further limitation arises from the symbolic outputs produced by automatic score transcription. Although all 128 MIDI pitches and predefined note-value categories are formally valid, transcription models typically do not impose explicit constraints from key, harmony, or meter. As a result, local pitch fluctuations, ornaments, or transcription errors may be converted into unnecessary chromatic notes, uncommon note values, or overly fragmented note sequences. Such audio-centric transcriptions can therefore be more detailed and less compositionally concise than intent-centric scores written by composers, potentially causing a train--inference mismatch.

\section{Conclusion}

We presented VocalRender, a score-native singing voice synthesis system that generates high-fidelity singing directly from composer-oriented symbolic scores. By combining an interleaved lyric--note representation with autoregressive diffusion modeling, VocalRender jointly determines expressive timing and output length without explicit duration prediction or time-aligned acoustic guidance. We also constructed a large-scale singing dataset with fine-grained score annotations to support scalable end-to-end training. Experiments demonstrate that VocalRender achieves strong intelligibility, speaker similarity, score-following ability, and perceptual naturalness across both in-domain and out-of-domain evaluations. Future work will focus on reducing the mismatch between automatically transcribed and composer-authored scores and incorporating stronger musical priors into symbolic score modeling.

\clearpage
\bibliography{aaai2027}

\clearpage
\appendix
\setcounter{secnumdepth}{1}

\section{Subjective Evaluation}

We recruit more than 20 participants for subjective evaluation, including students with music education backgrounds and amateurs. For PS-CMOS and MS-MOS test, only the qualified participants are allowed to take part, to ensure the musical information is evaluated precisely. For N-CMOS test, we further include amateurs to collect feedback from a more diverse audience. All participants are compensated for \$10 per hour. For each test, 20 samples or sample pairs are randomly selected from a pool of 1736 segments, without cherry-picking. The descriptions of the tests are listed below, followed by screenshots of the evaluation interfaces.

\subsection{Naturalness (N-CMOS)}
\begin{itemize}
    \item \textbf{System Interface:} Users listen to two singing samples, A and B, to compare their naturalness.
    \item \textbf{Questionnaire:} Which singing sample sounds more natural and human-like, rather than synthetic or AI-generated?
    \item \textbf{Evaluation Criteria:} Options include A+2 (Sample A is much more natural), A+1 (Sample A is slightly more natural), Tie (Both are equally natural), B+1 (Sample B is slightly more natural), and B+2 (Sample B is much more natural).
\end{itemize}

\subsection{Prosody Similarity (PS-CMOS)}
\begin{itemize}
    \item \textbf{System Interface:} Users listen to two singing samples, A and B, to evaluate their similarity to the reference prosody.
    \item \textbf{Questionnaire:} Ignoring vocal characteristics (timbre, style, vocal techniques), decide which sample's prosody (rhythm and melody) is more consistent with the reference.
    \item \textbf{Evaluation Criteria:} Options include A+2 (Sample A is much more similar), A+1 (Sample A is slightly more similar), Tie (Both are equally similar), B+1 (Sample B is slightly more similar), and B+2 (Sample B is much more similar).
\end{itemize}

\subsection{Music Score (MS-MOS)}

\begin{itemize}
    \item \textbf{System Interface:} Users listen to one singing sample while viewing its music score, to evaluate whether the singing follows the music score.
    \item \textbf{Questionnaire:} Does the singing follow the music score? A score melody player is also provided: it plays the notes of the score directly (a plain synthesized melody), to help you hear the intended tune. It is a reference only — do not rate its voice quality.
    \item \textbf{Evaluation Criteria:} Options include 1 (Unable to follow), 2 (Hardly follow), 3 (Roughly follow), 4 (Strictly follow).
\end{itemize}

\begin{figure}[htp]
\centering
\includegraphics[width=\columnwidth]{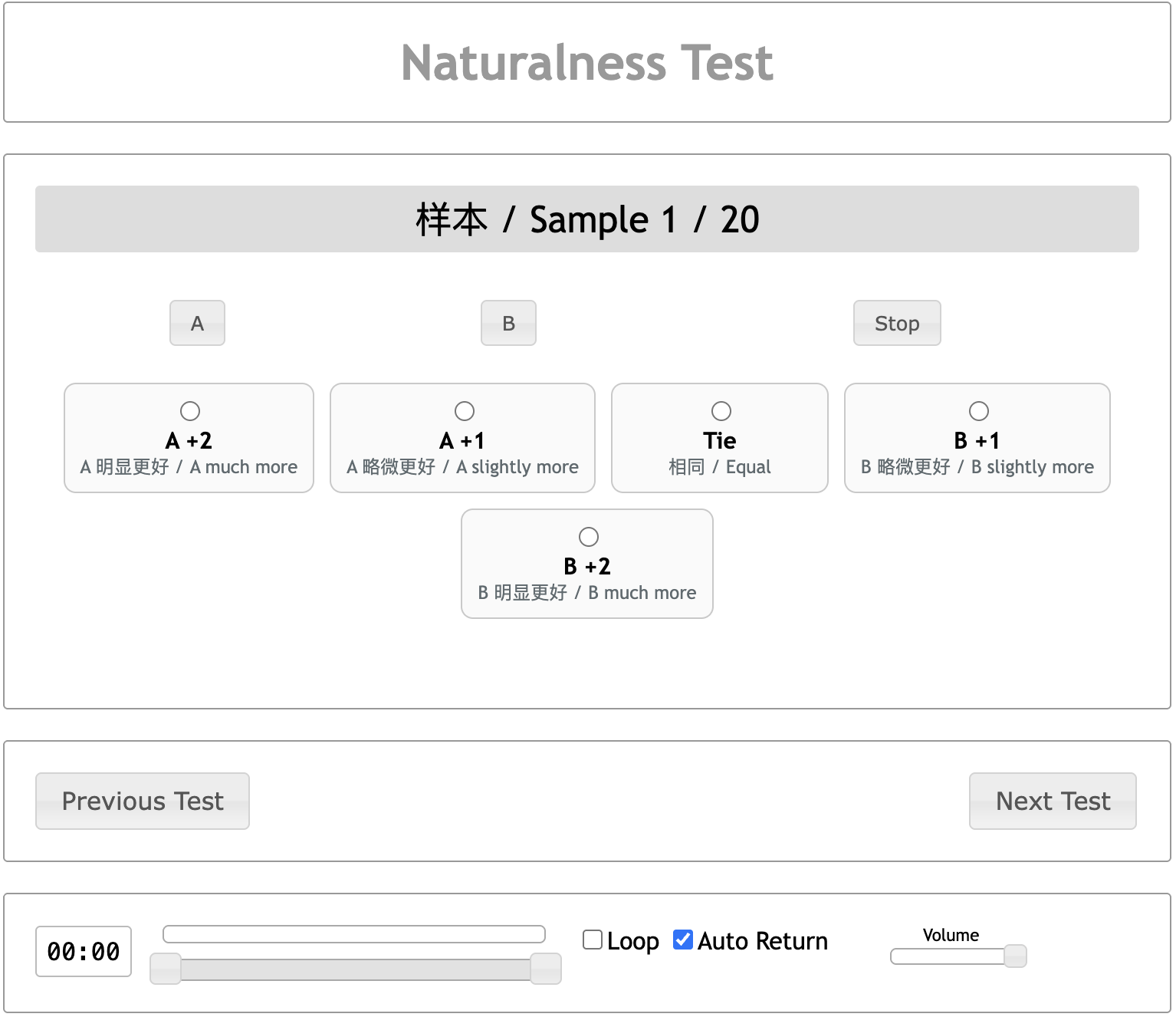}
\label{fig:N-CMOS}
\caption{Screenshot of N-CMOS test.}
\end{figure}

\begin{figure}[htp]
\centering
\includegraphics[width=\columnwidth]{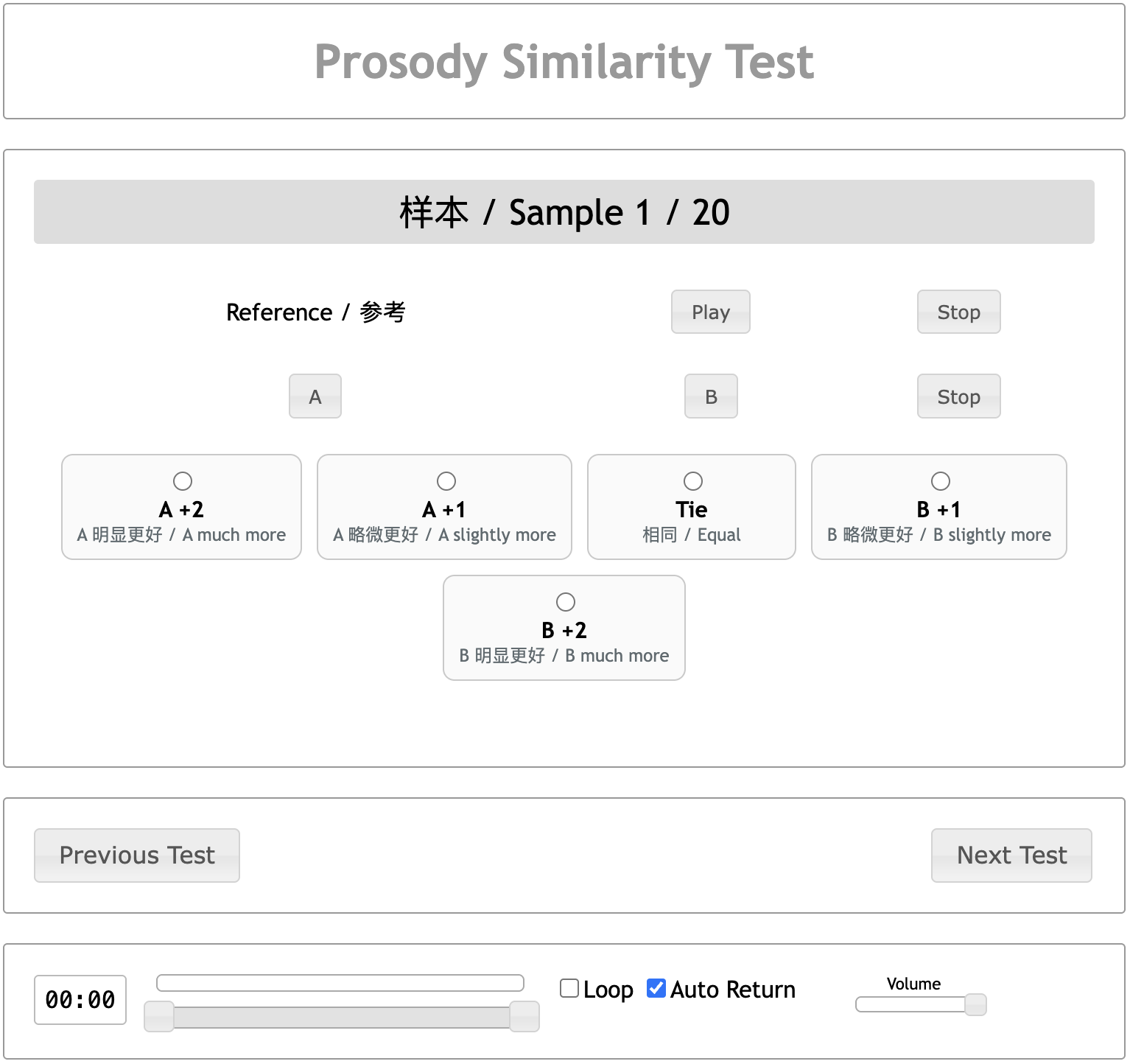}
\label{fig:PS-CMOS}
\caption{Screenshot of PS-CMOS test.}
\end{figure}

\begin{figure}[htp]
\centering
\includegraphics[width=\columnwidth]{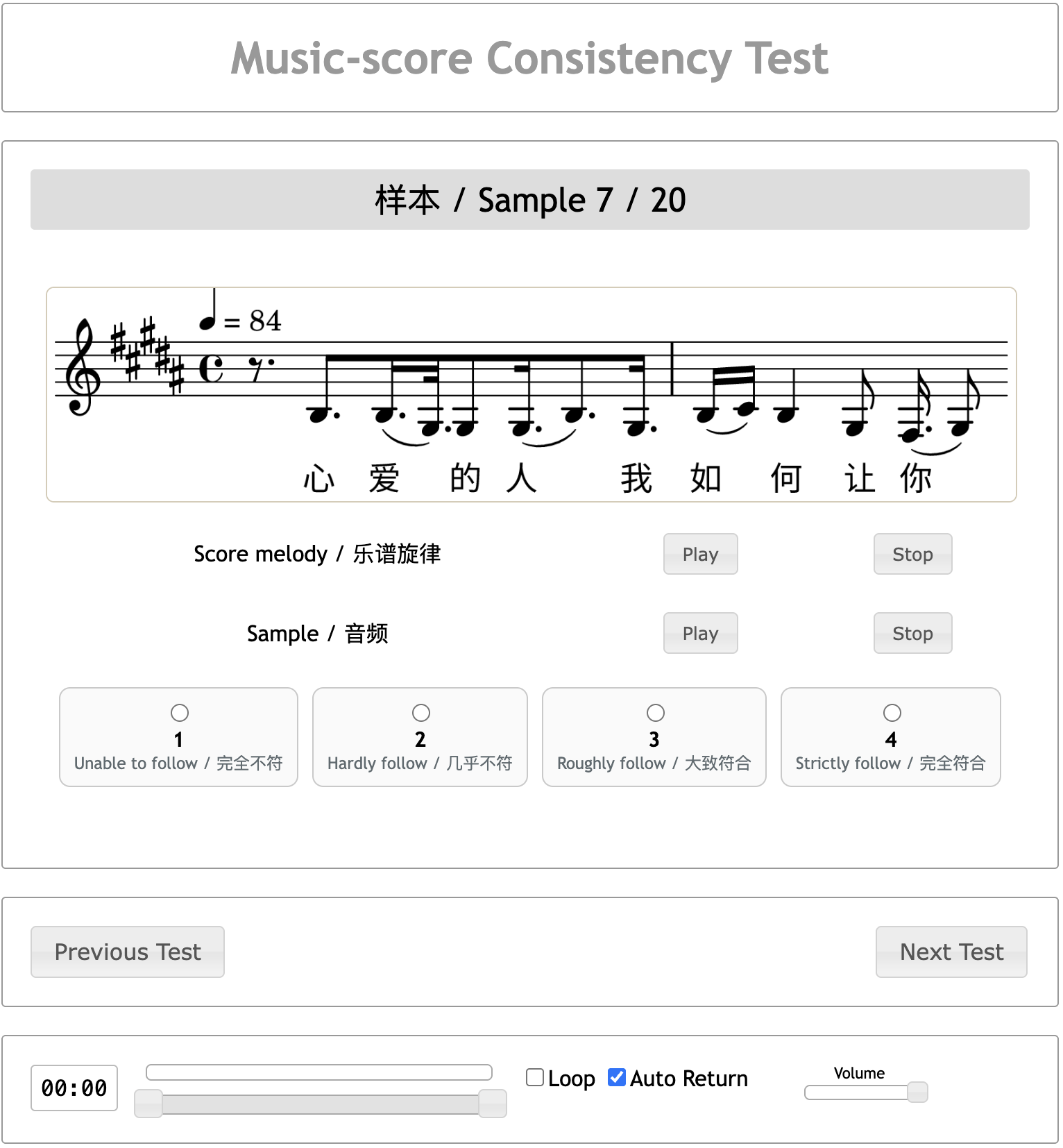}
\label{fig:MS-MOS}
\caption{Screenshot of MS-MOS test.}
\end{figure}

\section{CrawlSinger-OS}

\subsection{Data Construction}

\begin{figure}[htp]
\centering
\includegraphics[width=\columnwidth]{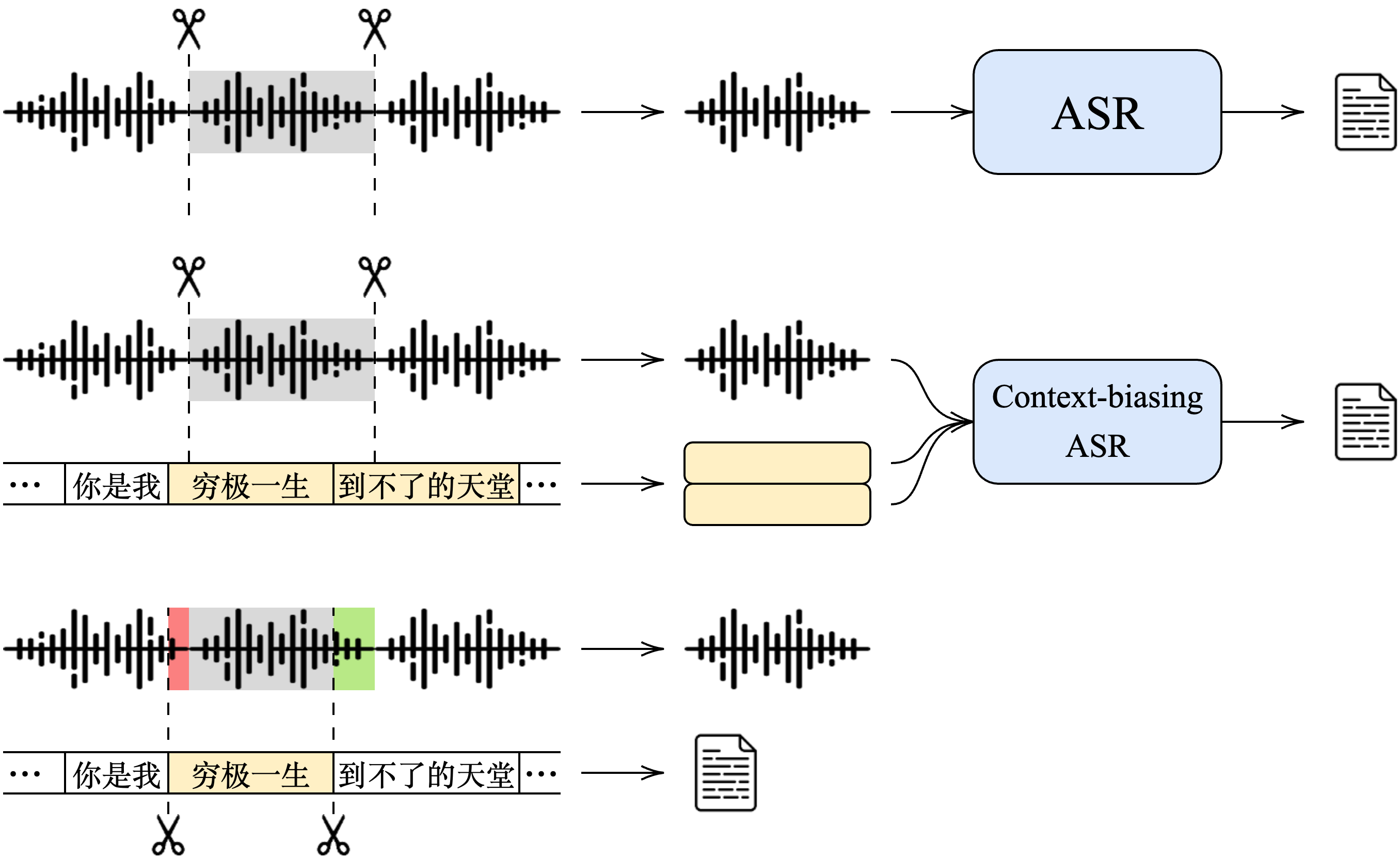}
\caption{Three processing branches for slicing and lyric transcription.}
\label{fig:transcribe_branches}
\end{figure}

We adopt SingCrawl pipeline \citep{chen2026vocalparse} to process data from multiple open-source datasets, including Muse, MuChin, SongFormDB and OpenSinger. The pipeline consists of four steps:

\begin{enumerate}
\item \textbf{Vocal Extraction}

Two mel-RoFormer-based models \citep{wang2023mel} are used in a cascaded manner to remove the background music and reverberation in the raw song audio, resulting in clean singing audio. We choose {\def\UrlFont{\ttfamily}\path{big_beta6x.ckpt}} and {\def\UrlFont{\ttfamily}\path{dereverb_mel_band_roformer_anvuew_sdr_19.1729.ckpt}} respectively for their validated performance.

\item \textbf{Slicing \& Lyric Transcription}

Most of SVS models still focus on sentence-level synthesis instead of whole song in one pass, so we slice the separated singing audio into segments less than 30 seconds, with paired lyric transcriptions. Considering different format and confidence of existing annotations in the source dataset, we design three branches of processing paths from weak label to strong label, as shown in Figure~\ref{fig:transcribe_branches}. \textit{Direct} path is designed for datasets with no available annotations. The slicing points are decided by the audio when silent region is detected, with an algorithm based on audio-slicer\footnote{\url{https://github.com/openvpi/audio-slicer}}. Then, the segments are fed into Qwen3-ASR \citep{shi2026qwen3} for lyric transcription. \textit{Candidate} path is designed for datasets with sentence-level lyric annotation, while the timestamps are not accurate enough for slicing (eg., from LRC file). The slicing step is the same as \textit{Direct} path, where a segment is given a start/end timestamp according to silent regions. Meanwhile, the lyrics from existing annotations will be chosen as candidate transcriptions when their regions are overlapped with specific segment. As the ideal transcription is likely to be a subset of candidate transcriptions, they are combined with the audio segment for context-biasing ASR. \textit{Refined} path is designed for datasets with relatively accurate sentence-level lyric annotation, while some off-cuts still exist if depending on its timestamps (eg., from YRC file). Unlike the previous two paths, \textit{Refined} path chooses slicing points according to annotations instead of waveform initially. Then, the initial slicing points are allowed to move to somewhere adjacent, based on naive voicing/non-voicing of RMVPE \citep{wei2023rmvpe}. The transcription of the segment follows the content of annotation, without involvement of ASR model.

\item \textbf{Force Alignment}

We retrain SOFA\footnote{\url{https://github.com/qiuqiao/SOFA}} on CrawlSinger and CrawlSinger-OS for a better alignment accuracy and stability. The original dictionary G2P is replaced by G2PW \citep{chen2022g2pw} for more reliable phoneme transform and unified processing for multiple datasets. Valided on a subset of GTSinger \citep{zhang2024gtsinger} and M4Singer \citep{zhang2022m4singer}, the SOFA weight retrained on large-scale data achieves Mean\_IOU of $0.84$ and VlablerEditRatio(50ms) of $0.092$.

\item \textbf{Pitch Transcription}

We use the official implementation of ROSVOT \citep{li2024robust} for pitch transcription, which achieves a RPA of $87.6$ reported in the origin paper. The pitch durations are further quantized into musical representation of note sequence and a global speed reference through algorithm proposed in SingCrawl.
\end{enumerate}

\subsection{Source Datasets}

\begin{enumerate}
\item \textbf{Muse}

Muse \citep{jiang2026muse} is a large-scale synthetic dataset, with approximately 7771 hours of songs generated by SunoV5. Based on the dataset, the authors successfully train a song generation model with competitive performance to those trained on real songs, indicating that the generated songs achieve competitive quality compared with human-performed recordings. We use the Chinese subset of Muse and implement step 1-4, with \textit{Refined} path of step 2, as the time-aligned lyrics have been provided by the generation system. Muse is released under the MIT License. as declared by the dataset authors.

\item \textbf{MuChin}

MuChin \citep{wang2024muchin} is a music description dataset with 6066 songs and time-aligned lyrics. We use full set of MuChin and implement step 1-4, with \textit{Candidate} path of step 2, as the provided timestamps are not reliable enough after manual inspection. The associated GitHub repository is released under the MIT License, while the song audio remains subject to the copyrights of the respective rights holders and the usage is limited to academic purposes only.

\item \textbf{SongFormDB}

SongFormDB \citep{hao2025songformer} is a music structure analysis corpus. We use the Ext subset with 4314 songs, and implement step 1-4, with \textit{Direct} path of step 2, as it contains no lyrics annotation. SongFormDB is released under License of CC BY 4.0 as declared by the dataset authors.

\item \textbf{OpenSinger}

OpenSinger \citep{huang2021multi} is a singing voice dataset of 53 hours, recording from 41 females and 25 males singers. Without word-level alignment and pitch-related annotations, OpenSinger is rarely considered for SVS training. As a result, we use full set of OpenSinger and implement step 3-4. OpenSinger is released under CC BY-NC-SA as declared by the dataset authors.
\end{enumerate}

The statistics of the processed source datasets, including duration, number of segments, audio type, and available annotations, are summarized in Table~\ref{tab:source_datasets}.

\subsection{Data Statistics}

\begin{table*}[t]
\centering
\begin{tabular}{lccccc}
\toprule
Source Dataset & Duration (hour) & Segments & Real/Synth & Audio Type & Annotation \\
\midrule
Muse & 2013 & 601k & Synth & Song & Lyric \\
MuChin & 158 & 84k & Real & Song & Lyric \\
SongFormDB & 93 & 48k & Real & Song & None \\
OpenSinger & 53 & 43k & Real & Singing & Lyric \\
\bottomrule
\end{tabular}
\caption{Statistics of source datasets after processing in CrawlSinger-OS.}
\label{tab:source_datasets}
\end{table*}

\begin{figure}[htp]
\centering
\includegraphics[width=\columnwidth]{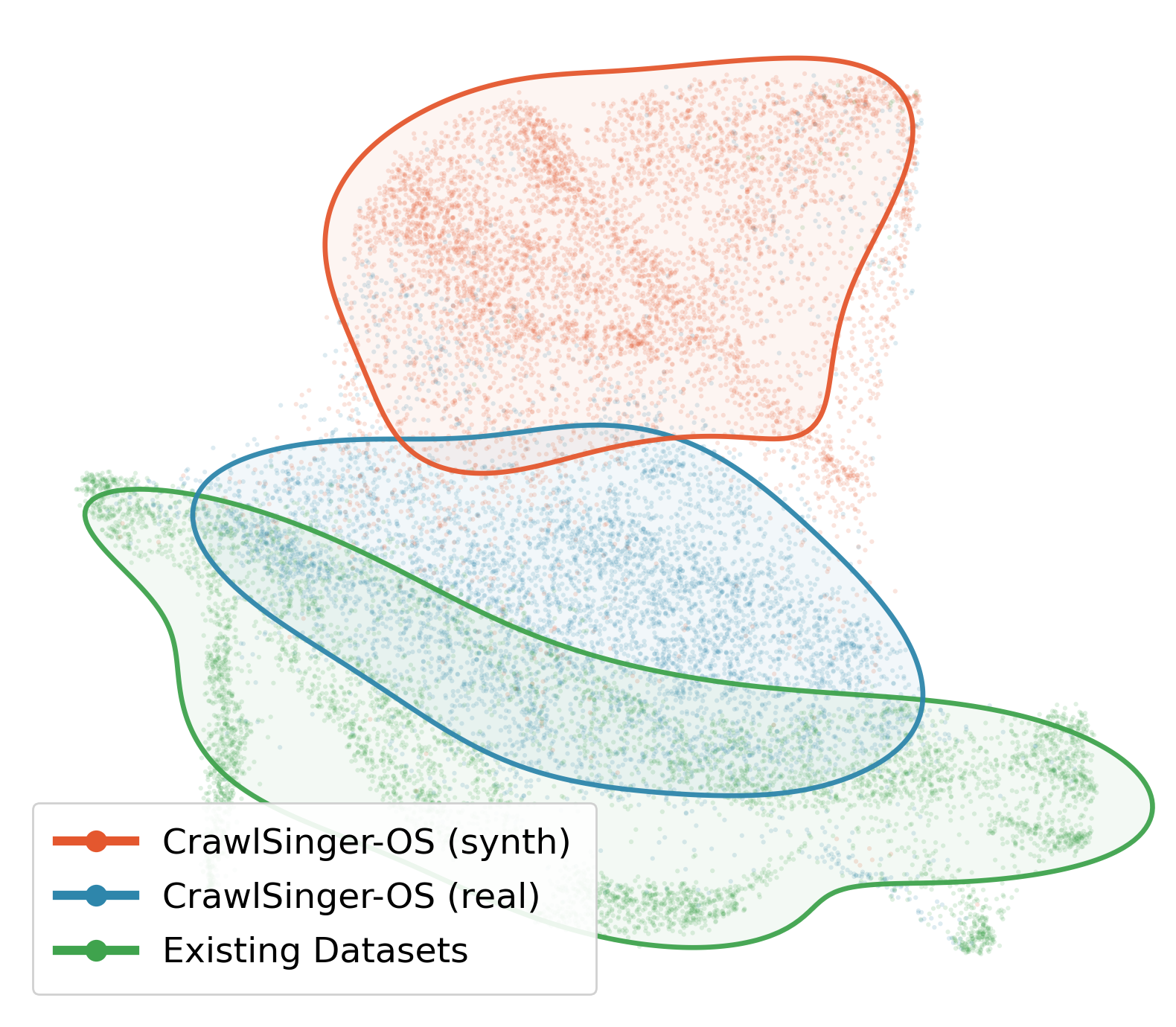}
\caption{Musical Diversity}
\label{fig:diversity_musical}
\end{figure}

\begin{figure}[htp]
\centering
\includegraphics[width=\columnwidth]{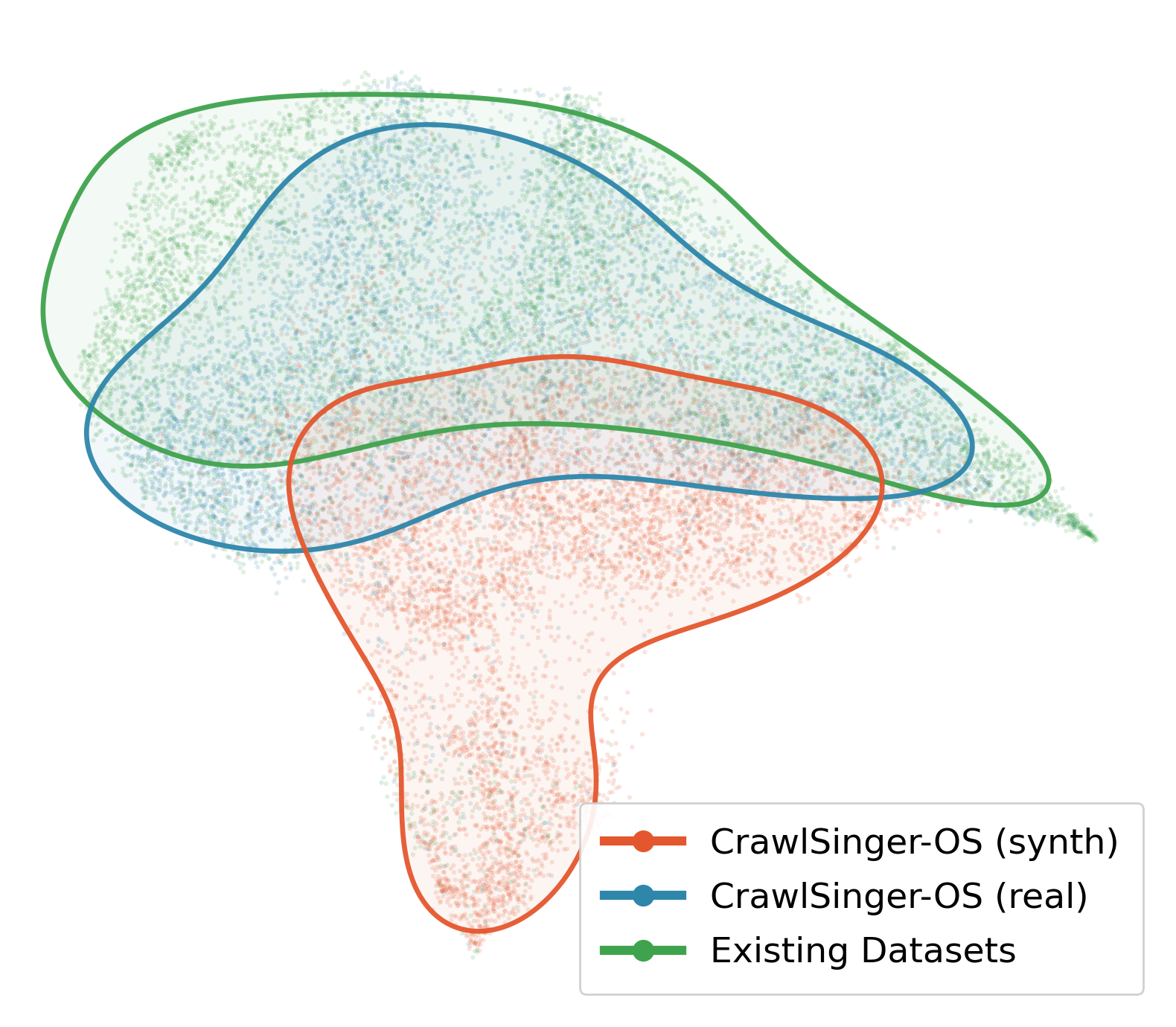}
\caption{Semantic Diversity}
\label{fig:diversity_semantic}
\end{figure}

\begin{figure*}[htp]
\centering
\includegraphics[width=\textwidth]{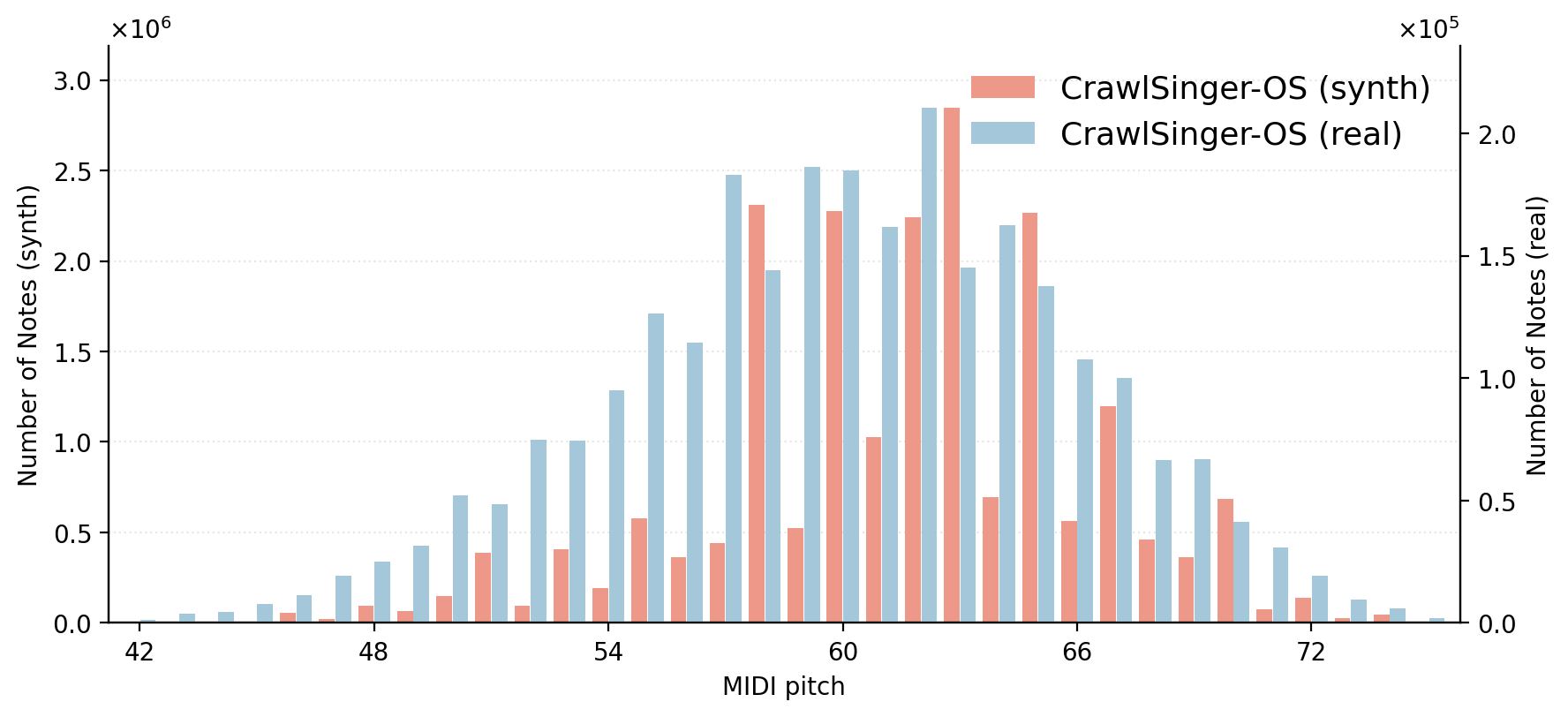}
\caption{Pitch Distribution of CrawlSinger-OS}
\label{fig:pitch_distribution}
\end{figure*}

To examine whether CrawlSinger-OS covers material that existing open singing corpora do not, we compare three pools: CrawlSinger-OS (synth), CrawlSinger-OS (real), and Existing Datasets (M4Singer + GTSinger). We sample 20,000 clips at random from each pool (60,000 in total). Each clip is described in two complementary spaces: a \emph{musical} embedding, taken as the time-averaged layer-24 hidden states of MERT-v1-330M\footnote{\url{https://huggingface.co/m-a-p/MERT-v1-330M}} \citep{li2024mert}, which encodes timbre, pitch and singing style; and a \emph{semantic} embedding, taken as the time-averaged last-layer hidden states of W2v-BERT 2.0\footnote{\url{https://huggingface.co/facebook/w2v-bert-2.0}} \citep{barrault2023seamlessm4t}, which encodes lyrical content. Both are standardized, reduced to 50 principal components, and projected to two dimensions with UMAP \citep{mcinnes2018umap} (\texttt{n\_neighbors=60}, \texttt{min\_dist=0.15}). Figure~\ref{fig:diversity_musical} and Figure~\ref{fig:diversity_semantic} show the resulting scatter. It can be seen that the synthetic data deviates from the real singing both musically and semantically, which suggests it should be used combined with real dataset to avoid potential distribution shifting. In contrast, CrawlSinger-OS (real) is complementary to the existing corpora such as GTSinger and M4Singer, as it covers exclusive region in musical dimensions. The CrawlSinger-OS (real) is largely overlapped with existing corpora in semantic dimensions, which hint they may contain similar song groups. CrawlSinger-OS (synth) also indicates a different pitch distribution with CrawlSinger-OS (real) as shown in Figure~\ref{fig:pitch_distribution}, where the pitches concentrate at specific values like $58,60,62,63,65$ rather than a smooth distribution.

\section{Additional Implementation Details}

\subsection{Training Configuration}

We initialize the model from the speech-pretrained VoxCPM2 \citep{zhou2026voxcpm2} checkpoint. The training uses the AdamW optimizer with DeepSpeed ZeRO-3 distributed optimization. For VocalRender-Pro, we train the model for 160k steps with a global batch size of 32,768 continuous tokens. The learning rate linearly increases to $1\times10^{-4}$ during the first 5k steps and then follows an inverse square-root decay schedule. The training speed is approximately 40k steps per day on 4 Nvidia H100 GPUs, resulting in a total training time of about 4 days. VocalRender adopts a two-stage training strategy. It is first pretrained on the synthetic subset for 40k steps and then finetuned on the realistic subset for 20k steps. During finetuning, the learning rate is reduced by half compared with pretraining. The complete training process takes approximately 1.5 days under the same hardware setting. For timbre-conditioned generation, we randomly sample a non-overlapping segment from the same song as the reference prompt, with a duration uniformly sampled between 2 and 8 seconds. All experiments are conducted with mixed-precision training using bfloat16.

\subsection{Model Scale and Generation Configuration}

The AR transformer contains approximately 1.7B parameters, while the DiT module contains approximately 0.6B parameters. The acoustic latent sequence is divided into patches with a patch size of 4 latent frames. Therefore, the effective language-model token rate is 6.25 Hz, derived from the 25 Hz latent frame rate. The VAE operates on 16 kHz input waveforms and reconstructs waveforms at 48 kHz. During inference, the DiT module uses 10 diffusion sampling steps with a classifier-free guidance scale of 2.0.

\end{document}